\documentclass[showpacs,prl,aps,superscriptaddress,twocolumn,floatfix]{revtex4}
\usepackage{graphicx,float,amsmath,amssymb}
\newfont{\ensmathquatorze}{msbm10 scaled 1400}
\newfont{\ensmathonze}{msbm10 scaled 1100}
\newfont{\ensmathdix}{msbm10}
\newfont{\ensmathneuf}{msbm10 scaled 833}
\newfont{\ensmathhuit}{msbm10 scaled 694}
\newfam\ensmathfam                        
\textfont\ensmathfam=\ensmathonze        
\scriptfont\ensmathfam=\ensmathdix       
\scriptscriptfont\ensmathfam=\ensmathhuit

\def\ensmf{\fam\ensmathfam\ensmathonze}         

\def\RR{{\ensmf R}}                 

\begin{document}

\title{Superfluidity versus Anderson localization in a dilute Bose gas}

\author{T. Paul}
\affiliation{Laboratoire de Physique Th\'eorique
et Mod\`eles Statistiques, CNRS,
Universit\'e Paris Sud, UMR8626, 91405 Orsay Cedex, France}
\author{P. Schlagheck}
\affiliation{Institut f{\"u}r Theoretische Physik,
Universit{\"a}t Regensburg, 93040 Regensburg, Germany}
\author{P. Leboeuf}
\author{N. Pavloff}
\affiliation{Laboratoire de Physique Th\'eorique
et Mod\`eles Statistiques, CNRS, 
Universit\'e Paris Sud, UMR8626, 91405 Orsay Cedex, France}
\begin{abstract}
We consider the motion of a quasi one dimensional beam of Bose-Einstein
condensed particles in a disordered region of finite extent. Interaction
effects lead to the appearance of two distinct regions of stationary flow. One
is subsonic and corresponds to superfluid motion. The other one is supersonic,
dissipative and shows Anderson localization. We compute analytically the
interaction-dependent localization length. We also explain the disappearance
of the supersonic stationary flow for large disordered samples.
\end{abstract}


\maketitle


Interference effects have a strong influence on the transport properties of
phase coherent systems. In particular in one dimension (1D) in presence of
disorder, they lead to Anderson localization (AL), revealed for instance by a
transmission decreasing exponentially with system size. On
the other hand, interactions in a low temperature Bose system may lead to
superfluidity (SF), i.e., perfect transmission. Understanding the interplay
between these two contrasting tendencies is one of the key issues of the
physics of interacting disordered Bose systems. Recent advances in creating
and manipulating guided cold atomic vapors and atom lasers (see, e.g., Refs.
\cite{gal,fragmen}) offers new prospects for studying such transport
phenomena in Bose-Einstein condensates (BEC). BEC systems are of particular
interest in that respect because they are almost perfectly phase coherent, and
interaction between their components can be easily modified and modeled.
Recently, great experimental efforts have been devoted to the identification
of AL in these systems \cite{Lye05,Cle05,Sch05}. Though not yet observed, the
possibility of AL has been pointed out theoretically in the non interacting
regime \cite{Gav05,SP06}, while interactions may play a role in the localization
scenario \cite{SP06}.

In this Letter we study a quasi 1D, weakly interacting BEC,
propagating through a disordered potential. In this context, localization has
been theoretically studied mainly for effective {\it attractive} interactions
(see, e.g., \cite{Gre92} and references therein), with less attention on the
{\it repulsive} case that we consider here (see, however, Refs.
\cite{Bil05,Pau05,Bil06}). In the regime of weak disorder and weak velocity,
the condensate depletion and the destruction of SF have been considered
theoretically in Refs. \cite{weakdis1}, while the strong disorder regime has
been analyzed numerically in Ref. \cite{weakdis2}.

\begin{figure}
\includegraphics*[width=\columnwidth]{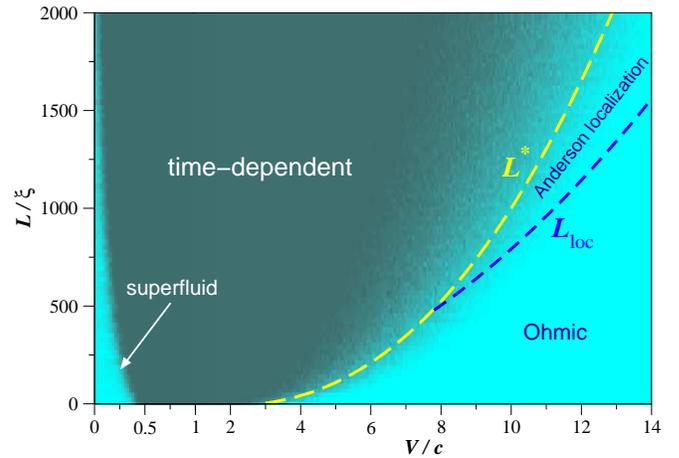}
\caption{(color online) Transport of a quasi 1D BEC with normalized velocity
$v$ through a disordered potential of length $L$ [Eq.~(\ref{e2}),
$\lambda=n_{\rm i} \xi = 0.5$]. Dark region: time dependent flow; light
blue/gray regions: stationary flow. Note the enlarged scale for $v\in[0,1]$.
}\label{f1}
\end{figure}

Our main result is a new, global picture of coherent transport of an
interacting BEC through disorder, with a clear characterization of the
different transport regimes where SF and AL can be observed. Our findings are
summarized in Fig.~\ref{f1}. The different types of flow are displayed as a
function of $v = V/c$ and $L/\xi$, where $V$ is the condensate velocity
relative to a disordered potential of spatial extension $L$, $c$ is the speed
of sound, and $\xi$ the healing length. At low velocities (subsonic regime),
the flow is SF, as expected on the basis of Landau's criterion. The density
profile is stationary, but locally modified by the external potential at the
expense of the creation of a normal fluid fraction, that we compute
perturbatively, extending previous results obtained in the limit $v
\rightarrow 0$ \cite{weakdis1} to finite velocities [Eq. (\ref{e4})]. In the
opposite supersonic regime, a region of stationary flows also exists, but in
this case energy dissipation occurs. In this domain, depending on the extent
of the disordered potential, the system is either in an ohmic or in an AL
regime (respectively characterized by a transmission decreasing linearly 
or exponentially with $L$, see below). In between the high and low velocity
stationary regimes there is an intermediate domain, centered around $v \sim
1$, where the flow is time dependent.

Consider a BEC at rest in a 3D guide where a disordered potential $U$ of finite
spatial extension $L$ is moved in the longitudinal $x$ direction at constant
velocity $V>0$. The transverse confinement is such that the ``1D mean field
regime'' \cite{Men02} holds. The system is then described by an order
parameter of the form $\exp\{-i\mu t/\hbar\}\psi(x,t)$, which is solution of
\begin{equation}\label{e1}
i\,\hbar\,\frac{\partial\psi}{\partial t}=-\frac{\hbar^2}{2
	m}\frac{\partial^2\psi}{\partial x^2} + \left[
U(x-Vt)+g\,|\psi|^2-\mu\right]\psi\; .
\end{equation}
In the case of particles experiencing an effective repulsive interaction
characterized by the 3D s-wave scattering length $a>0$, the constant $g$ reads
$g=2h \nu_\perp a$, where $\nu_\perp$ is the frequency of the
transverse harmonic trap \cite{Ols98}. In the stationary regime, where the
flow is time--independent in the frame moving with the potential, $\psi$
depends on $x$ and $t$ only through the variable $X=x-Vt$. The appropriate
boundary condition is $\psi(X\to-\infty)=\sqrt{n_0}$ 
(where $n_0$ is a constant) \cite{Leb01}. The condensate is characterized
by a chemical potential $\mu=g\, n_0$, a speed of sound $c=(\mu/m)^{1/2}$ and
a healing length $\xi=\hbar/(m\,c)$. We consider a regime where the typical
value of the disordered potential is much smaller than $\mu + m V^2 /2$. This
regime is easily reached experimentally and is important for our purpose
because it corresponds to a range of parameters where AL is not blurred by
effects connected to ``fragmentation of the condensate'' (see, e.g.,
Ref.~\cite{fragmen} and references therein).

The picture we draw in the present Letter is generic and does not
depend on the specific structure of the disordered potential $U$. We
illustrate our findings with a potential of the form
\begin{equation}\label{e2}
U(x)=\lambda\,\mu\,\xi\sum_{n=1}^{N_{\rm i}}\delta(x-x_n) \; ,
\end{equation}
where the $x_n$ are the random, uncorrelated positions of $N_{\rm i}$
 delta scatterers uniformly distributed over a length $L$ (with a density 
$n_{\rm i}=N_{\rm i}/L$). $\lambda>0$ is the
dimensionless strength of a scatterer. This potential has a mean value
$\langle U\rangle =\lambda\,\mu\,\xi\,n_{\rm i}$ and a cumulant
$\langle U(x)U(0)\rangle-\langle U\rangle^2=(\hbar^2/m)^2 \sigma\,\delta(x)$,
where $\sigma=\lambda^2\,n_{\rm i}\,\xi^{-2}$. From what is known in the case of
Schr\"odinger equation, this potential is typical insofar as
localization properties are concerned.

Numerically, we
solve Eq.~(\ref{e1}) using the potential (\ref{e2}) 
for given $\mu$, $m$, $n_{\rm i}$, $\lambda$ and $V$. We consider 100
realizations of the random potential. The fraction $f$ of potentials for which
a stationary solution exists is plotted in Fig.~\ref{f1} using a gray scale
(dark, $f=0$; light blue/gray, $f=1$) as a function of the normalized
variables $L/\xi$ and $v=V/c$. This normalization rescales all interaction
effects. In the following we characterize the different stationary regimes 
represented in Fig.~\ref{f1}.

Let us start with the subsonic stationary regime. There,
the density profile is easily computed
perturbatively to be of the form $n(X)=n_0+\delta n(X)$, with
\cite{Hak97,Leb01} 
\begin{equation}\label{e3}
\delta n(X)\simeq -\frac{m\, n_0}{\hbar^2\kappa}
\int_{\scriptscriptstyle \RR} {\rm d}y \, U(y)\, \exp\{-2\, \kappa |X-y|\} \; ,
\end{equation}
where 
\begin{equation}\label{kappa}
\kappa=\frac{m}{\hbar}\left|c^2-V^2\right|^{1/2}
\end{equation}
is an effective wave vector. From Eq.~(\ref{e3}) $\delta n (X\to \pm \infty) =
0$ and the density is only perturbed in the region of the potential. This
corresponds to perfect transmission of the condensate through the disordered
potential. There is no energy dissipation nor drag exerted on the potential, a
characteristic feature of SF \cite{Pav02}. We have verified these
properties numerically and also checked the stability of the SF flow. There is
also a non superfluid fraction, since a momentum $P=\hbar\int{\rm d}x\, \delta
n\, \partial_x S$ \cite{Bar93} can be associated to the flow (where $S$ is the
phase of the order parameter). In the spirit of Landau's approach for
determining the normal fraction in liquid $^4$He (see also \cite{Ast04}), one
can associate to this momentum a normal mass $M_{\rm n}=P/V$. In the
perturbative limit where Eq. (\ref{e3}) holds one gets
\begin{equation}\label{e4}
\frac{ M_{\rm n}}{M} =  \frac{m^2}{2\,\hbar^4\,\kappa^3\,L} 
\int_{{\scriptscriptstyle\RR}^2} 
{\rm d}y_1{\rm d}y_2 U(y_1)U(y_2)\, K(y_2-y_1)\; ,
\end{equation}
where $K(y)=(1+2\kappa|y|)\exp\{-2\kappa |y|\}$ and $M=m\,n_0\,L$ is the mass
of a region of size $L$ of the unperturbed condensate. The ratio $M_{\rm n}/M$
corresponds to the normal fraction present in the disordered region. For a
potential of type (\ref{e2}), at zero velocity, in the scarce impurities limit
($1\gg n_{\rm i}\xi$), Eq.~(\ref{e4}) yields $M_{\rm n}(0)/M=\lambda^2 n_{\rm
i}\xi/2$, in accordance with the findings of Refs. \cite{weakdis1,Ast04}. The
present approach allows to extend this result to finite velocities, yielding
(for $\kappa\gg n_{\rm i}$)~: $M_{\rm n}(v)/M_{\rm n}(0)=(1-v^2)^{-3/2}$.

\begin{figure}
\includegraphics*[width=\columnwidth,height=0.7\columnwidth]{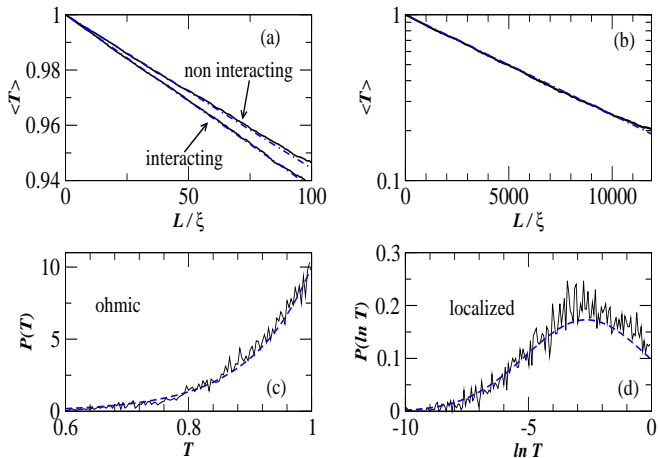}
\caption{Statistical properties of the transmission $T$ at $v=30$ for
potential (\ref{e2}) with $\lambda=n_{\rm i}\xi=0.5$ [except plot (a) where
$v=3$ and $\lambda=0.1$]. Dashed line: analytical result; solid line:
numerics. Top row: averaged transmission. Analytical and numerical results are
almost indistinguishable. The non interacting results are also represented in
plot (a). Bottom row:  Probability distribution of $T$. Case (c): ohmic
regime, $L/L_{\rm loc}\simeq 0.1$; case (d): AL regime, $L/L_{\rm loc}\simeq
2.4$.}\label{f2}
\end{figure}

The situation changes drastically as the velocity increases. Starting
from the SF subsonic region (with fixed $U$ and $L$), at some critical
velocity the system enters a domain where the flow becomes unsteady (dark region
in Fig.~\ref{f1}). The velocity at which this transition occurs depends very
much on the specific form of the potential $U$ (see, e.g., the discussion of
the influence of the value of $\lambda$ for a regular array of equally spaced
impurities in \cite{Tar99}).
If $v$ further increases one gets into a new
stationary region (now supersonic). Within this region, if $v$ is large enough,
one reaches a domain (qualified as ohmic in Fig.~\ref{f1}) where
perturbation theory holds, yielding \cite{Leb01}
\begin{equation}\label{e6}
\delta n(X)\simeq \frac{2\,m\,n_0}{\hbar^2\,\kappa}
\int_{-\infty}^X\!\!\!\!{\rm d}y \, U(y)\,
 \sin[2\kappa(X-y)] \; ,
\end{equation}
where $\kappa$ is given by Eq.~(\ref{kappa}). In the region upstream with
respect to the disordered potential ($X \rightarrow + \infty$), Eq.~(\ref{e6})
corresponds to an oscillatory wave, i.e., to a non--perfect transmission,
indicating the dissipative nature of the flow. Equation (\ref{e6}) allows to
express the disorder averaged transmission coefficient as \cite{Pau05}~:
\begin{equation}\label{e7}
\langle T\rangle\simeq 1 - \frac{L}{L_{\rm loc}(\kappa)} \;\; , \quad
\mbox{where} \quad L_{\rm loc}(\kappa)=\frac{\kappa^2}{\sigma}
\; .
\end{equation}
The probability distribution of $T$ can be shown to be 
\begin{equation}\label{e7b}
P(T)=\frac{L_{\rm loc}}{L} \exp\left\{
-(1-T) \frac{L_{\rm loc}}{L} \right\}\; .
\end{equation}
The validity of Eqs.~(\ref{e7},\ref{e7b}) is confirmed by comparison with
numerical results in Figs.~\ref{f2}(a) and \ref{f2}(c). 
These equations hold
for potentials of the form (\ref{e2}), for Gaussian white noise potentials as
well as for correlated noise, but in the latter case the expression of $L_{\rm
loc}(\kappa)$ is modified. In this regime the
transmission decreases linearly with $L$, this is why we call it ohmic. 
Eqs. (\ref{e6},\ref{e7},\ref{e7b}) are valid in the
perturbative regime $|\delta n|\ll n_0$, or equivalently $L\ll L_{\rm
loc}(\kappa)$. Starting from the ohmic regime, keeping fixed $v>1$ and
increasing the length $L$ of the disordered potential, one enters the
interesting non-perturbative domain $L > L_{\rm loc}$ where AL (i.e., an
exponential suppression of the transmission with disorder length) is expected.
In this regime, we will show [cf. Eq. (\ref{e11}) below] that the quantity
$L_{\rm loc}$ appearing in Eq.~(\ref{e7}) corresponds indeed to the
localization length.

\begin{figure}
\includegraphics[width=\columnwidth]{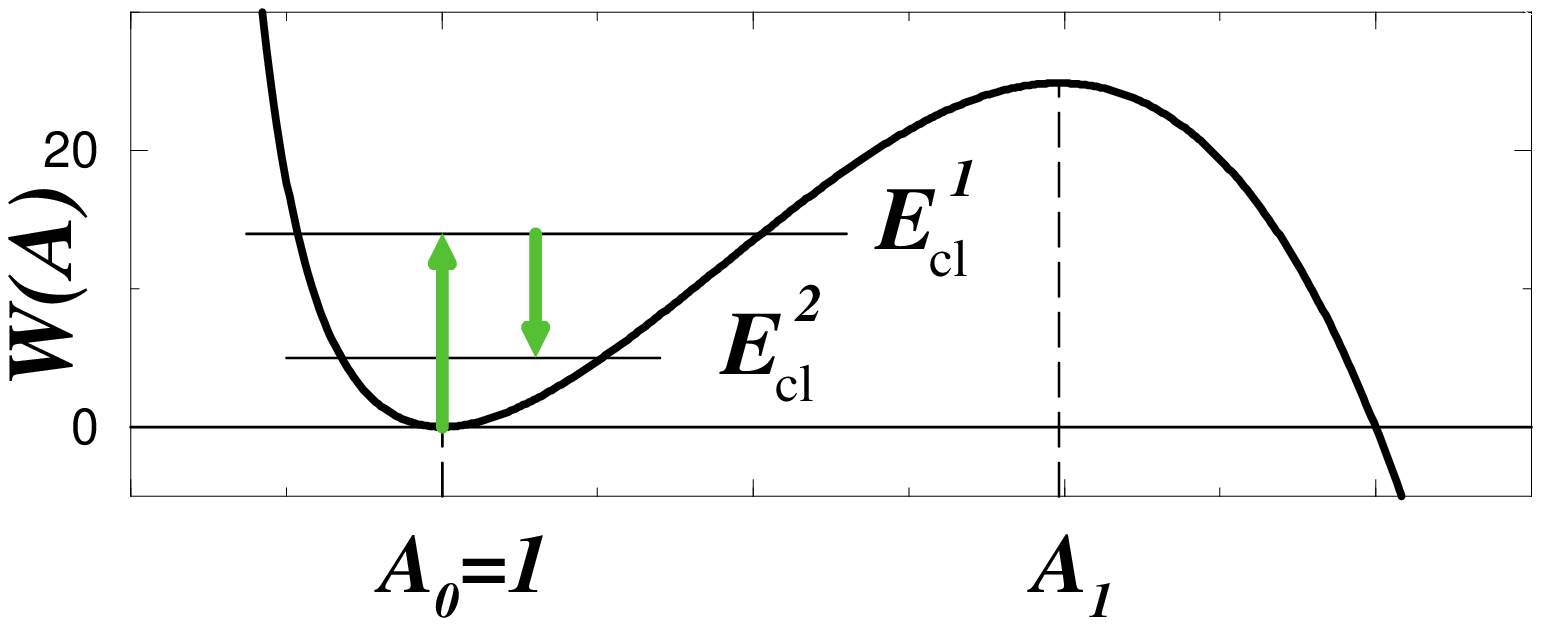}
\includegraphics[width=\columnwidth]{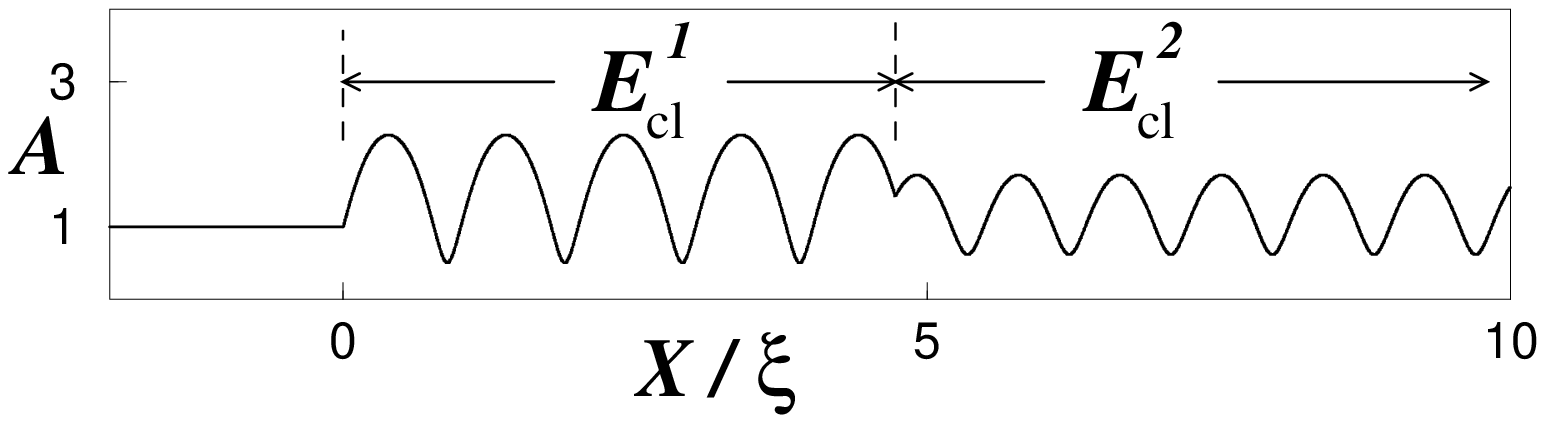}
\caption{Upper panel: $W$ as a function of $A=|\psi|/\sqrt{n_0}$ (drawn for
$v=4$). $A_0(=1)$ and $A_1$ are the zeros of ${\rm d}W/{\rm d}A$. The
fictitious particle is initially at rest with $E_{\rm cl}^0=0$. The value of
$E_{\rm cl}$ changes at each impurity. The lower panel displays the
corresponding oscillations of $A(X)$, with two impurities (vertical dashed
lines) at $x_1 = 0$ and $x_2 = 4.7 \, \xi$. }\label{f3}
\end{figure}

We now derive a non-perturbative treatment of the supersonic stationary flow,
valid in the AL regime. We consider a disordered potential of type (\ref{e2})
and look for stationary solutions. Between two impurities ($x_{n}$ and $x_{n+1}$ say) 
the random potential is zero and (\ref{e1}) admits a first
integral of the form \cite{Leb01}
\begin{equation}\label{e8}
\frac{\xi^2}{2}\left(\frac{{\rm d}A}{{\rm d}X}\right)^2 + W[A(X)]
=E^n_{\rm cl}
\; ,
\end{equation}
where $E^n_{\rm cl}$ is a constant, $A=|\psi|/\sqrt{n_0}$ and
$W(A)=\frac{1}{2}(A^2-1)(1+v^2-A^2-v^2/A^2)$.
Eq.~(\ref{e8}) has a Hamiltonian form, expressing energy conservation
for a fictitious classical particle with ``mass'' $\xi^2$, ``position'' $A$,
``time'' $X$, evolving in a potential $W$ (whose typical shape is displayed
in Fig.~\ref{f3}). The integration of Eq.~(\ref{e8}) starts from
the left, using the initial ``position'' $A(X<0)=1$, which corresponds to
$E_{\rm cl}^0=W(1)=0$ (supersonic uniform flow, see \cite{Leb01}). 
The behavior of $A$ for $X>L$ depends on the final value
of $E^{N_{\rm i}}_{\rm cl}$. A stationary solution exists only if $A(X>L)$
remains bounded, i.e., if $E^{N_{\rm i}}_{\rm cl}<W(A_1)$ (where $A_1$
corresponds to the local maximum of $W$, see Fig.~\ref{f3}). In this case, the
transmission coefficient is \cite{Leb03}
\begin{equation}\label{e10}
T=\frac{1}{\displaystyle1+(2\kappa^2\,\xi^2)^{-1}E^{N_{\rm i}}_{\rm cl}}
\; .
\end{equation}
This formula is non-perturbative in the deep supersonic regime $v \gg 1$ (
i.e., it is not limited by the condition $|\delta n|\ll n_0$)
and $T$, given by Eq.~(\ref{e10}), may become very small 
(contrarily to Eq.~(\ref{e7})
which is only valid for values of the transmission close to unity). 

In order to describe the density profile in the disordered region and to
determine the value of $E_{\rm cl}^{N_{\rm i}}$, one can device a diffusion
equation describing the evolution of $E_{\rm cl}$ under the effect of the
kicks caused by each randomly placed impurity (cf. Fig.~\ref{f3}). This allows
--through (\ref{e10})-- to compute the disorder average of the transmission
and its probability distribution. For $L\gg L_{\rm loc}$ and $\kappa \gg
n_{\rm i}, \xi^{-1}$ one obtains
\begin{equation}\label{e11}
\langle \ln T \rangle = - L / L_{\rm loc} (\kappa) \ ,
\end{equation}
where $L_{\rm loc}(\kappa)$ is given by Eq. (\ref{e7}), whereas the
probability distribution reads
\begin{equation}\label{e12}
P(\ln T)=\sqrt{\frac{L_{\rm loc}}{4\pi L}}
\exp\left\{-\frac{L_{\rm loc}}{4 L}
\left(\frac{L}{L_{\rm loc}}+\ln T\right)^2
\right\} \; .
\end{equation}
Excellent agreement of the analytical formulas (\ref{e11}) and (\ref{e12})
with the numerical results is found, cf. Figs.~\ref{f2}(b) and \ref{f2}(d). It
is interesting to note that the exponential decrease (\ref{e11}) of the mean
transmission and the log normal distribution (\ref{e12}) are characteristic
features of AL of linear waves with a localization length $L_{\rm loc}$ (see,
e.g., \cite{Tig99}). Hence, ``typical'' AL is predicted to occur in supersonic
interacting BEC systems. Interaction induces a modification
of the wave vector in $L_{\rm loc}$: the expression (\ref{e7}) for the
localization length coincides with the non-interacting one, but computed for
an effective interaction--dependent wave vector $\kappa$, instead of $k=m
V/\hbar$. The repulsive interaction diminishes the available kinetic energy,
and therefore reduces the localization length with respect to the
non-interacting case (since $\kappa < k$).

The present analysis also explains the onset of time dependence observed
numerically in the supersonic region at fixed $V$ and increasing $L$ (see 
Fig.~\ref{f1}). Once the average ``classical energy'' $\langle E_{\rm cl}^{N_{\rm
i}} \rangle$ exceeds the value $W(A_1)$ (see Fig.~\ref{f3}), the number of
stationary solutions decreases dramatically. This occurs for $L>L^*(\kappa)$
with
\begin{equation}\label{e13}
L^*(\kappa)=L_{\rm loc}(\kappa)\,\ln\left( v^2 /8 \right)/2 \; .
\end{equation}
This expression fits very well the numerically determined
frontier of the stationary supersonic domain (see Fig.~\ref{f1}).

At moderate supersonic velocities Eq.~(\ref{e13}) shows that 
$L^*$ is roughly of the same order as $L_{\rm loc}$ and one cannot reach 
the regime $L_{\rm loc}\ll L < L^*$
where the log normal distribution (\ref{e12}) is observed.
This is the reason why Fig.~\ref{f2}(b) displays the distribution at relatively
large velocity ($v=30$, $L^* /L_{\rm loc}\simeq 2.4$). However, the
exponential decay (\ref{e11}) of the average transmission is observed for {\it
all} stationary supersonic velocities, with $L_{\rm loc}$ given by
Eq.~(\ref{e7}), as confirmed by Figs.~\ref{f2}(a) and \ref{f2}(b).

The $v$--dependence of the two relevant length scales $L_{\rm loc}$ and $L^*$
is shown in Fig.~1. Supersonic stationary solutions exist above a critical
velocity $v_1$ corresponding to $L^*(\kappa_1)=0$. One gets $v_1
=\sqrt{8}\simeq 2.8$, which is the onset of the ohmic regime. In contrast, AL
exists only within the stationary regime for $L_{\rm loc}<L<L^*$ (cf.
Fig.~\ref{f1}). This is possible only if $v > v_2$, where $L^* (\kappa_2)=
L_{\rm loc}(\kappa_2)$, i.e., $v_2 = {\rm e} \sqrt{8}\simeq 7.7$. However,
for $v > v_2$, the difference of $L_{\rm loc}(\kappa)$ with its
non-interacting counterpart $L_{\rm loc}(k)$ is less than $2\%$. Hence
interactions play no significant role in the AL regime. They
will be relevant however in the low--velocity, small--di\-sor\-der--length
sector of the ohmic regime. At $v=3$ for instance, they induce a modification
of $11 \%$ of the localization length [cf. the comparison of interacting and
non interacting results in Fig.~\ref{f2}(a)].

In summary, we provide a complete picture of the quasi 1D transport of a
weakly interacting BEC through a disordered potential. We find that
interactions have several major and experimentally relevant consequences~:
(i) existence of a subsonic SF regime, (ii) existence of non steady flows
around $v\simeq 1$, (iii) renormalization of the localization length in the
stationary supersonic regime, and (iv) introduction of a maximum disorder
length scale $L^*$ at which AL disappears and time dependence sets in. Explicit
methods to characterize experimentally the different regimes through, e.g., 
the heating rate, will be discussed elsewhere.

This work was supported by grants ANR--05--Nano--008--02 and
ANR--NT05--2--42103, by the IFRAF Institute and by the Alexander von Humboldt
Foundation.


\begin{thebibliography}{99}
\bibitem{gal} A. Greiner {\it et al.}, cond-mat/0701328;
W. Guerin {\it et al.},  Phys. Rev. Lett. {\bf 97}, 200402 
(2006); 
P. Hommelhoff, W. H\"ansel, T. Steinmetz, T. W. H\"ansch and J.
Reichel, New J. Phys. {\bf 7}, 3 (2005); 
T. Lahaye {\it et al.}, Phys. Rev. Lett. {\bf 93}, 093003 (2004); 
R. Folman {\it et al.}, Adv. At.
Mol. Opt. Phys. {\bf 48}, 263 (2002);  
\bibitem{fragmen} J. Fort\'agh and C. Zimmermann, Rev. Mod. Phys. {\bf 79},
235 (2007). 
\bibitem{Lye05} J. E. Lye {\it et al.}, Phys. Rev. Lett. {\bf 95}, 070401
(2005); C. Fort et al., Phys. Rev. Lett. {\bf 95}, 170410 (2005). 
\bibitem{Cle05} D. Cl\'ement {\it et al.}, Phys. Rev. Lett. {\bf 95}, 170409
(2005); D. Clement {\it et al.}, New J Phys. {\bf 8} 165
(2006). 
\bibitem{Sch05} T. Schulte {\it et al.}, Phys. Rev. Lett. {\bf 95}, 170411
	(2005).
\bibitem{Gav05} U. Gavish and Y. Castin, Phys. Rev. Lett. {\bf 95}, 020401
	(2005); R. C. Kuhn {\it et al.}, Phys. Rev. Lett. {\bf 95}, 250403 (2005);
P. Massignan and Y. Castin Phys. Rev. A {\bf 74}, 013616 (2006).
\bibitem{SP06} L. Sanchez-Palencia {\it et al.}, cond-mat/0612670;
B. Shapiro, cond-mat/070134.
\bibitem{Gre92} S. A. Gredeskul and Y. S. Kivshar, Phys. Rep. {\bf 216}, 1
(1992).
\bibitem{Bil05} N. Bilas and N. Pavloff, Phys. Rev. Lett. {\bf 95}, 130403
(2005).
\bibitem{Pau05} T. Paul, P. Leboeuf, N. Pavloff, K. Richter, and P.
	Schlagheck, Phys. Rev. A {\bf 72},  063621 (2005).
\bibitem{Bil06} N. Bilas and N. Pavloff, Eur. Phys. J. D {\bf 40}, 387 (2006).
\bibitem{weakdis1} K. Huang and H. F. Meng, Phys. Rev. Lett. {\bf 69}, 644
	(1992); S. Giorgini, L. Pitaevskii, and S. Stringari, Phys. Rev. B {\bf 49},
	12938 (1994).
\bibitem{weakdis2} G. E. Astrakharchik, J. Boronat, J. Casulleras, and S.
	Giorgini, Phys. Rev. A {\bf 66}, 023603 (2002).
\bibitem{Men02} C. Menotti and S. Stringari, Phys. Rev. A {\bf 66},
043610 (2002).
\bibitem{Ols98} M. Olshanii, Phys. Rev. Lett. {\bf 81}, 938 (1998).
\bibitem{Leb01} P. Leboeuf and N. Pavloff, Phys. Rev. A {\bf 64}, 033602 (2001).
\bibitem{Hak97} V. Hakim, Phys. Rev. E {\bf 55}, 2835 (1997).
\bibitem{Pav02} N. Pavloff, Phys. Rev. A {\bf 66}, 013610 (2002). 
\bibitem{Bar93} I. V. Barashenkov and E. Yu. Panova, Physica D {\bf 69}, 114
(1993).
\bibitem{Ast04} G. E. Astrakharchik and L. P. Pitaevskii, Phys. Rev. A {\bf
	70}, 013608 (2004).
\bibitem{Tar99} D. Taras-Semchuk and J. M. F. Gunn, Phys. Rev. B {\bf 60},
13139 (1999).
\bibitem{Leb03} P. Leboeuf, N. Pavloff, and S. Sinha, Phys. Rev. A {\bf 68}, 
063608 (2003).
\bibitem{Tig99} B. A. Van Tiggelen in {\it Diffusive Waves in Complex Media},
	J. P. Fouque ed., p. 1 (Kluwer Academic Publishers, Dordrecht, 1999).
\end{thebibliography}
\end{document}